\documentclass{sig-alternate-05-2015}

\usepackage{amsmath,amssymb,epsfig}
\usepackage{array}
\usepackage{booktabs}
\usepackage{graphicx}
\usepackage{epstopdf}
\usepackage{times}
\usepackage{color}

\newcommand{\eat}[1]{}

\begin{document}

\makeatletter
\def\@copyrightspace{\relax}
\makeatother

\title{Hybrid Deep-Semantic Matrix Factorization for Tag-Aware Personalized Recommendation}

\author{%
\alignauthor
{\fontsize{12}{12}
Zhenghua Xu$^{\footnotemark[1]}$  \quad Cheng Chen$^{\footnotemark[2]}$  \quad Thomas Lukasiewicz$^{\footnotemark[1]}$ \quad  Yishu Miao$^{\footnotemark[1]}$}%
\vspace{2mm}\\
{\fontsize{10}{10}
$^{\footnotemark[1]}$ Department of Computer Science, University of Oxford, United Kingdom\\
\fontsize{10}{10}
$\{$zhenghua.xu, thomas.lukasiewicz, yishu.miao$\}$@cs.ox.ac.uk}\vspace{-0.6mm}\\
\fontsize{10}{10}
$^{\footnotemark[2]}$ School of Computer Science, Beijing University of Posts and Telecommunications, China\\
\fontsize{10}{10}
ccbupt@bupt.edu.cn\\
}

\maketitle

\begin{abstract}
Matrix factorization has now become a dominant solution for personalized recommendation on the Social Web. To alleviate the cold start problem, previous approaches have incorporated various additional sources of information into  traditional matrix factorization models. These upgraded models, however, achieve only ``marginal'' enhancements on the performance of personalized recommendation. Therefore, inspired by the recent development of deep-semantic modeling, we propose a hybrid deep-semantic matrix factorization (HDMF) model to further improve the performance of tag-aware personalized recommendation by integrating the techniques of deep-semantic modeling, hybrid learning, and matrix factorization. Experimental results show that HDMF significantly outperforms the state-of-the-art baselines in tag-aware personalized recommendation, in terms of all evaluation metrics, e.g., its mean reciprocal rank (resp., mean average precision) is $1.52$ (resp., $1.66$) times as high as that of the best baseline.

\end{abstract}



\vspace*{-1ex}

\section{Introduction}
\label{sec:Introduction}

In 
the Web 2.0, social tagging systems are introduced by many websites, where users can freely annotate online items using arbitrary tags (commonly known as \emph{folksonomy}~\cite{Hotho06}). Since social tags are good summaries of the relevant items and the users' preferences, and since they also contain little sensitive information about their creators, they are valuable information for privacy-enhanced personalized recommendation.
Consequently, many efforts have been put on tag-aware personalized recommendation using \emph{content-based filtering}~\cite{cantador2010content, Shepitsen:2008,Xu2016} or \emph{collaborative filtering}~\cite{BouadjenekHB13,Ma2011,Mano2016,Tso-Sutter:2008}.

However, as users can freely choose their own vocabulary, social tags may contain many uncontrolled vocabularies. This usually results in sparse, redundant, and ambiguous tag information, and significantly weakens the performance of content-based recommendation systems. The common solution is to apply machine learning techniques, e.g., clustering~\cite{Shepitsen:2008} or {autoencoders}~\cite{Zuo2016}, to learn more abstract and representative features from raw tags. Recently, Xu et al.~\cite{Xu2016} propose a deep-semantic model called DSPR which utilizes deep neural networks to model abstract and recommendation-oriented representations for social tags. DSPR is reported to achieve better performance than the clustering and autoencoder solutions.

\emph{Matrix factorization} is a collaborative-filtering-based solution, which has become a dominant solution for personalized recommendation on the Social Web~\cite{BouadjenekHB13,Ma2011,Mano2016} and has been reported to be superior to memory-based techniques~\cite{koren2009}.
However, there exists a \emph{cold start} problem in matrix factorization:
many users only give very few ratings, resulting in a very sparse user-item rating matrix, and making it difficult to summarize users' preferences.
A widely adopted solution is to incorporate additional sources of information about users, e.g., implicit feedback~\cite{koren2009}, social friendship~\cite{Ma2011}, geographical neighborhood~\cite{hu2014your}, or textual comments~\cite{Mano2016}. We call these upgraded models \emph{additional-information-based matrix factorization (AMF)} models.

Although the existing deep-semantic model, DSPR, and the upgraded matrix factorization models, AMF, have progressively improved the tag-aware personalized recommendation, there are still a few drawbacks:
(i) DSPR does not utilize the idea of collaborative filtering; hence, the valuable correlation information between users and items is not being used to help recommendation.
(ii) As a deep neural model, DSPR stacks many layers, which makes it difficult to optimize the model by gradient back-propagation.
(iii) The existing AMF models generally incorporate the additional information as a regularization term of matrix factorization; this term's coefficient, as proved in~\cite{Ma2011}, has to be very small; therefore, the additional information has very limited contribution on the optimizing gradient, resulting in only ``marginal'' improvements on the recommendation performance.
(iv) The recommendation results of the existing AMF models are difficult to interpret, because latent factor matrices are used to represent users and items.

Consequently, to solve the above problems and to further improve the performance of tag-aware personalized recommendation, we propose a \emph{hybrid deep-semantic matrix factorization (HDMF)} mo\-del, which integrates the techniques of deep-semantic modeling, hybrid learning, and matrix factorization. Generally, HDMF uses a \emph{tag-based user matrix} and a \emph{tag-based item matrix} as respective inputs of two deep autoencoders to generate \emph{deep-semantic user and item matrices} at the code layers, and also \emph{reconstructed user and item matrices} at the output layers. The deep model is then trained by using a \emph{hybrid learning signal} to minimize both \emph{reconstruction errors} and \emph{deep-semantic matrix factorization errors}, i.e., the squared differences between the user-item rating matrix (seeing tags as positive ratings) and the dot product of deep-semantic user and item matrices (seeing deep-semantic matrices as the decomposed matrices in matrix factorization). The intuitions of using the hybrid learning signal are: (i) minimizing reconstruction errors can learn better representations for both users and items; (ii) deep-semantic matrix factorization offers a learning signal that connects users and items to discover the underlying users' preferences; (iii) two signals can complement each other to provide sufficient gradients for better model optimization and escaping the local minima.

HDMF thus has the following advantages.
(i) It overcomes the drawback of DSPR by adding collaborative-based capabilities to the deep-semantic model.
(ii) The hybrid learning signal helps HD\-MF to better optimize the model and escape local minima.
(iii)~Differently from AMF models, the additional tag information in HD\-MF is directly used to model the decomposed user and item matrices in matrix factorization; this thus maximizes the effect of the additional tag information on the model optimization.
(iv) HDMF remedies the non-interpretability problem in matrix factorization:
considering deep-semantic matrices as the decomposed matrices and finding the most influential input tags for each dimension, the decomposed user and item matrices in HDMF become interpretable.


The main contributions of this paper are briefly as follows:
\begin{itemize}
\vspace{-1.5ex}\item We briefly analyze the state-of-the-art personalized recommendation models that use content-based filtering or matrix factorization and identify their existing problems.
\vspace{-1.7ex}\item We innovatively propose a hybrid deep-semantic matrix factorization (HDMF) model to tackle these problems and to further improve the performance of tag-aware personalized recommendation, by integrating the techniques of deep-seman\-tic modeling, hybrid learning, and matrix factorization.
\vspace{-1.7ex}\item Experimental results show that HDMF significantly outperforms the state-of-the-art baselines in tag-aware personalized recommendation, in terms of all evaluation metrics, e.g., its mean reciprocal rank (resp., mean average precision) is $1.52$ (resp., $1.66$) times as high as that of the best baseline.


\end{itemize}

\vspace*{-1.5ex}

\section{Preliminaries}
\label{sec:preliminary}

A \emph{folksonomy} is a tuple $\mathcal{F}=(U,T,D,A)$, where $U$, $T$, and $D$ are sets of {\em users}, {\em tags}, and {\em items}, respectively, and $A \subseteq U\times T\times D$ is a set of assignments $(u,t,d)$ of tag $t$ to item $d$ by user $u$~\cite{Hotho06}.

A \emph{tag-based user profile} is a feature vector $x=[g^u_{i}]^{|T|}_{i=1}$, where $|T|$ is the tag vocabulary's size, and $g^u_{i}=|\{(u,t_i,d)\in A\mid d \,{\in}\, D\}|$ is the number of times that user $u$ annotates items with tag~$t_i$; the \emph{tag-based user matrix} is thus defined as $X=[x_i]^{|U|}_{i=1}$, where $x_i$ is the profile vector of the $i${th} user, and $|U|$ is the total number of users. Similarly, a \emph{tag-based item profile} is a vector $y=[g^d_{j}]^{|T|}_{j=1}$, where $g^d_{j}=|\{(u,t_j,d)\,{\in}\, A\mid u \,{\in}\, U\}|$ is the number of times that item $d$ is annotated with tag~$t_j$; while the \emph{tag-based item matrix} is defined as $Y=[y_j]^{|D|}_{j=1}$, where $y_j$ is the profile vector of the $j$th item, and $|D|$ is the total number of items.


The \emph{user-item rating matrix} is $R=[r_{i,j}]^{|U|,|D|}_{i=1,j=1}$, where $r_{i,j}$ is the number of tags annotated by user $i$ to item $j$. Given $R$,  traditional \emph{matrix-facto\-rization-based recommender systems} aim to approximate $R$ using the decomposed latent matrices of users and items, i.e., $X^l$ and $Y^l$, respectively, which are optimized by minimizing the squared differences between $R$ and ${X^l}^T \cdot Y^l$ on a set of observed ratings; formally,

\vspace{-1.2em}
\begin{small}
\begin{align}\label{equ:4}
 \min_{X^l,Y^l} \sum_{i=1}^{|U|}\sum_{j=1}^{|D|}I_{i,j}(r_{i,j}-{x_i^l}^T\cdot y_j^l)^2,
\end{align}
\end{small}
\vspace{-1em}

\noindent where $I_{i,j}$ is $1$, if user $i$ annotated item $j$, and $0$, otherwise~\cite{Mano2016}. After optimization learning, the \emph{predicted user-item rating matrix} $\hat{R}={X^l}^T \cdot Y^l$ is used for personalized recommendation.

\vspace*{-1.5ex}

\section{Hybrid Deep-semantic Matrix\protect\newline Factorization}
\label{sec:Deep-Semantic-Similarity-Model}

To alleviate the cold start problem in traditional matrix factorization, a widely adopted solution is to incorporate additional sources of information about users to achieve additional-information-based matrix factorization (AMF)~\cite{hu2014your,koren2009,Ma2011,Mano2016}. However, as analyzed in Section~\ref{sec:Introduction} and demonstrated by both our experimental results and the results reported in~\cite{Mano2016}, the existing AMF models achieve only ``marginal'' (around $5\%$ in~\cite{Mano2016}) improvements on the performance of personalized recommendation.
Therefore, inspired by the recent development of deep-semantic modeling~\cite{Xu2016}, we propose a hybrid deep-semantic matrix factorization (HDMF) model to tackle these problems and to further enhance the performance of tag-aware personalized recommendation, by integrating the techniques of deep-semantic modeling, hybrid learning, and matrix factorization.

\begin{small}
\begin{figure}[!t]
\vspace{-0.5em}
\begin{center}
      \epsfig{figure=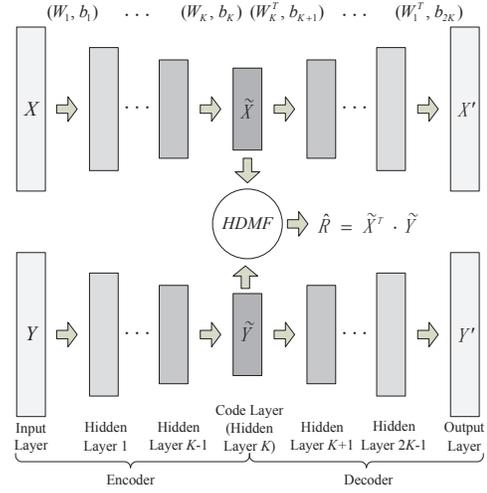,angle=0, width=2.5in}
\end{center}
\vspace{-2em}
\caption{Overview of HDMF}
\vspace{-1.5em}
\label{fig:overview}
\end{figure}
\end{small}

Figure~\ref{fig:overview} shows an overview of the HDMF model. Generally, HDMF takes the tag-based user and item matrices $X$ and $Y$ (defined in Section~\ref{sec:preliminary}) as inputs of two deep autoencoders, consisting of \emph{encoders} and \emph{decoders}. These inputs are then passed through multiple hidden layers and projected to the deep-semantic user and item matrices $\widetilde X$ and $\widetilde Y$ at the code layers, and to the reconstructed user and item matrices $X'$ and $Y'$ at the output layers. The HDMF model is then trained by using a hybrid learning signal to minimize both deep-semantic matrix factorization errors and reconstruction errors. Finally, a predicted user-item rating matrix $\hat{R}={\widetilde X}^T \cdot \widetilde Y$ is used for personalized recommendation.

\subsection{\!Deep-Semantic Matrix Factorization}

Deep-semantic matrix factorization is solely based on the encoder parts of the deep autoencoders, which can be seen as multi-layer perception networks.
Formally, given the tag-based user and item matrices $X$ and $Y$, a weight matrix $W_1$, and a bias vector~$b_1$, the intermediate outputs $h_1(\cdot)$ of the first hidden layers in the encoders are defined as follows:

\vspace{-1.2em}
\begin{small}
\begin{align}
h_1(X)=\tanh(W_1 X+b_1), \qquad
h_1(Y)=\tanh(W_1 Y+b_1),
\end{align}\label{equ:1}
\end{small}
\vspace{-1.5em}

\noindent where $\tanh$ is used as the activation function. Similarly, the intermediate outputs of the $j${th} hidden layers $h_j(\cdot)$, $j\in \{2,\ldots,K\}$, in the encoders are defined as follows:

\vspace{-1.2em}
\begin{small}
\begin{align}\label{equ:2}
h_j(X)&=\tanh(W_jh_{j-1}(X)+b_j), \\
h_j(Y)&=\tanh(W_jh_{j-1}(Y)+b_j),
\end{align}
\end{small}
\vspace{-1.5em}

\noindent where $W_j$ and $b_j$ are the weight matrix and the bias vector for the $j${th} hidden layers in the encoders, respectively, and $K$ is the total number of hidden layers in each encoder.

Then, the outputs of the $K${th} hidden layers, i.e., the code layers, are the deep-semantic user and item matrices, denoted $\widetilde X$ and $\widetilde Y$, respectively. Formally,

\vspace{-1.2em}
\begin{small}
\begin{align}\label{equ:3}
\widetilde X=h_K(X), \qquad \widetilde Y=h_K(Y).
\end{align}
\end{small}
\vspace{-1.5em}

Consequently, by seeing the deep-semantic matrices $\widetilde X$ and $\widetilde Y$ as the decomposed user and item matrices in matrix factorization, the parameters $W_j$ and $b_j$ can be optimized by minimizing the following deep-semantic matrix factorization errors:

\vspace{-1em}
\begin{small}
\begin{align}\label{equ:6}
\hspace{-1em}
L_{DMF}(\Theta)=&\,(1-\lambda_\theta)\sum_{i=1}^{|U|}\sum_{j=1}^{|D|}I_{i,j}(r_{i,j}-{\widetilde x_i}^T\cdot \widetilde y_j)^2 \nonumber\\
&+\,\lambda_\theta(\sum\limits_{j=1}^{K}\|W_j\|^2+\sum\limits_{j=1}^{K}\|b_j\|^2),
\end{align}
\end{small}
\vspace{-0.75em}

\noindent where $r_{i,j}$ is an element in the user-item rating matrix $R$, indicating the number of tags assigned by user $i$ to item $j$; \,$\widetilde x_i$ (resp., $\widetilde y_j$) is the vector at the $i$th (resp., $j$th) column of $\widetilde X$ (resp., $\widetilde Y$), which is the deep-semantic representation of the $i$th user (resp., $j$th item); the second term is a regularization term used to prevent overfitting, and $\lambda_\theta$ is the regularization parameter.

\subsection{Hybrid Learning Signal}
However, it is difficult to train the model using solely the learning signal from deep-semantic matrix factorization.
This is because the model stacks many layers of non-linearities, and when learning signals are back-propa\-gated to the first few layers, they become minuscule and insignificant to learn good representations for the users and items, which in turn results in poor local minima.
A common solution is to first pre-train each layer using restricted Boltzmann machines (RBMs) \cite{hinton2006fast,hinton2006reducing} or autoencoders~\cite{bengio2007greedy} and then use back-propaga\-tion to fine-tune the entire deep neural network~\cite{erhan2010does}.

Therefore, in this work, we directly incorporate autoencoders into the deep-semantic matrix factorization model, and train the deep model using a hybrid learning signal that integrates reconstruction errors of autoencoders with the deep-semantic matrix factorization errors. We thus call this model hybrid deep-semantic matrix factorization (HDMF). The intuition behind it 
is as follows: (i) the reconstruction-error-based signal can learn better representations for both users and items; (ii) the collaborative learning signal from deep-semantic matrix factorization can connect users and items to discover the underlying users' preferences; (iii) furthermore, the reconstruction-error-based signal can complement deep-semantic matrix factorization to provide sufficient gradients for better optimizing the model and escaping the local minima.





Specifically, as shown in Figure~\ref{fig:overview}, we adopt autoencoders with tied weights in HDMF, i.e., the weight matrices in the decoder are the transposes of weight matrices in the encoder. The decoders take the deep-semantic user and item matrices $\widetilde X$ and $\widetilde Y$ at the code layer as the inputs and generate reconstructed user and item matrices $X'$ and $Y'$ at their output layers. Then, reconstruction errors are computed based on the squared differences between the original tag-based matrices ($X$ and $Y$) and the reconstructed matrices ($X'$ and $Y'$). Finally, the reconstruction-error-based learning signal will be used to first update $W^{T}_1$, then back-propagated to update $W^{T}_2$, $W^{T}_3$, and so on. As updating $W^T_j$ is equivalent to updating $W_j$, this signal complements  deep-semantic matrix factorization and offers sufficient gradients to the first few layers of the deep model.




Formally, the intermediate outputs of the $K{+}j$th hidden layers $h_{K+j}(\cdot)$, $j\in \{1,\ldots,K-1\}$, in the decoders are defined as:

\vspace{-0.95em}
\begin{small}
\begin{align}\label{equ:7}
h_{K+j}(X)&=\tanh(W^T_{K-(j-1)}h_{K+(j-1)}(X)+b_{K+j}), \\
h_{K+j}(Y)&=\tanh(W^T_{K-(j-1)}h_{K+(j-1)}(Y)+b_{K+j}),
\end{align}
\end{small}
\vspace{-0.95em}

\noindent where $W^T_{K-(j-1)}$ is the transpose of $W_{K-(j-1)}$, and $b_{K+j}$ is the bias vector for the $K{+}j$th hidden layer. The outputs of the $2K{-}1$th hidden layers are used to generate reconstructed user and item profiles, denoted $X'$ and $Y'$, at the output layers: 

\vspace{-1.2em}
\begin{small}
\begin{align}\label{equ:8}
X'&=\tanh(W^T_1h_{2K\!-\!1}(X)+b_{2K}),\\
Y'&=\tanh(W^T_1h_{2K\!-\!1}(Y)+b_{2K}).
\end{align}
\end{small}
\vspace{-1.2em}

Then, the reconstruction errors of the user (resp., item) matrix are computed as the sum of the Euclidean (i.e., L$2$) norms of the differences between the tag-based user (resp., item) profile $x_i$ (resp., $y_j$) in $X$ (resp., $Y$) and the reconstructed user (resp., item) profile $x'_i$ (resp., $y'_j$) in $X'$ (resp., $Y'$). By integrating the reconstruction errors with the deep-semantic matrix factorizations errors (as defined in Equation~\ref{equ:6}), the HDMF model is thus trained by minimizing the following hybrid learning signal:

\vspace{-2em}
\begin{small}
\begin{align}\label{equ:11}
\hspace{-1em}
L_{HDMF}(\Theta)=&\,(1-\lambda_\theta-\lambda_e)\sum_{i=1}^{|U|}\sum_{j=1}^{|D|}I_{i,j}(r_{i,j}-{\widetilde x_i}^T\cdot \widetilde y_j)^2 \nonumber\\
&+\,\lambda_e(\sum_{i=1}^{|U|} \|x'_i-x_i\|+\sum_{j=1}^{|D|}\|y'_{j}-y_{j}\|) \nonumber\\
&+\,\lambda_\theta(\sum\limits_{j=1}^{K}\|W_j\|^2+\sum\limits_{j=1}^{2K}\|b_j\|^2).
\end{align}
\end{small}
\vspace{-1em}

\noindent


\section{Experiments}
\label{sec:ExperimentalStudy}

We have conducted extensive experimental studies and compar\-ed our proposed hybrid deep-semantic matrix factorization (HD\-MF) model with a number of state-of-the-art baselines, which are grouped into two categories and summarized as follows:

\noindent\textbf{Content-based tag-aware models.} Four state-of-the-art models that utilize social tags as the content information to conduct tag-aware personalized recommendation are selected as the baselines. Similarly to HDMF, they all apply machine learning techniques to model abstract and effective representations for users or/and items; i.e., the clustering-based models, \textbf{CCS} and \textbf{CCF} \cite{Shepitsen:2008}, the auto\-encoder-based model, \textbf{ACF}~\cite{Zuo2016}, and the deep-seman\-tic similarity-based model, \textbf{DSPR}~\cite{Xu2016}. 

\noindent\textbf{Matrix-factorization-based models.} Three matrix-factorization-based recommendation models are also selected as the baselines; i.e., the traditional matrix factorization model, \textbf{MF}, and the addition\-al-information-based matrix factorization (AMF) models, \textbf{MF$_{sf}$} \cite{Ma2011} and \textbf{MF$_{tc}$} \cite{Mano2016}, which incorporate, respectively, the social friendships and the textual comments of users as the additional sources of information for matrix factorization.

\vspace*{2ex}

\begin{small}
\begin{table}[!t]
\vspace{-0.6em}
\caption{Dataset Information}%
\centering
\smallskip \begin{tabular}{cccc}
\toprule
Users ($u$) & Tags ($t$) & Items ($i$) & Assignments (($u,t,i$))\\\midrule
1\,843 & 3\,508 & 65\,877 & 339\,744\\\bottomrule
\end{tabular}\label{tab:dataset}
\vspace{-1em}
\end{table}
\end{small}

\begin{small}
\begin{table*}[!t]
\vspace{-0.5em}
    \caption{Recommendation Performance of Various Models (in $\%$)}
      \centering
\smallskip \begin{tabular}
{p{1.45cm}<{\centering}|
p{0.75cm}<{\centering}p{0.75cm}<{\centering}
p{0.75cm}<{\centering}p{0.75cm}<{\centering}|
p{0.75cm}<{\centering}p{0.75cm}<{\centering}
p{0.75cm}<{\centering}p{0.75cm}<{\centering}|
p{0.75cm}<{\centering}p{0.75cm}<{\centering}
p{0.75cm}<{\centering}p{0.75cm}<{\centering}|
p{0.75cm}<{\centering}|p{0.75cm}<{\centering}
}
\toprule
Models & $P@5$ & $P@15$ & $P@30$ & $P@50$ & $R@5$ & $R@15$ & $R@30$ & $R@50$ & $F@5$ & $F@15$ & $F@30$ & $F@50$ & $MAP$ & $MRR$\\
\midrule
CCF & $0.913$ & $0.757$ & $0.597$ & $0.454$ & $0.439$ & $1.051$ & $1.499$ & $1.803$ & $0.593$ & $0.880$ & $0.854$ & $0.726$ & $0.437$ & $0.200$\\
ACF & $1.120$ & $0.909$ & $0.736$ & $0.595$ & $0.590$ & $1.209$ & $1.917$ & $2.364$ & $0.791$ & $1.038$ & $1.064$ & $0.950$  & $0.637$ & $0.252$\\
CCS & $2.397$ & $1.903$ & $1.564$ & $1.273$ & $0.938$ & $2.271$ & $3.739$ & $4.774$ & $1.349$ & $2.070$ & $2.205$ & $2.010$ & $1.319$  & $0.523$ \\
DSPR & $13.34$& $9.285$ & $6.950$ & $5.306$ & $4.235$ & $8.347$ & $12.00$ & $14.98$ & $6.430$ & $8.791$ & $8.803$ & $7.836$ & $5.452$ & $2.547$ \\
\midrule
MF & $9.157$ & $7.467$ & $6.784$ & $6.302$ & $1.302$ & $2.851$ & $4.988$ & $7.587$ & $2.280$ & $4.127$ & $5.749$ & $6.899$ & $6.757$ & $1.682$\\
MF$_{sf}$ & $10.16$ & $8.063$ & $7.302$ & $6.736$ & $1.457$ & $3.109$ & $5.407$ & $8.132$ & $2.549$ & $4.487$ & $6.213$ & $7.368$ & $6.920$ & $1.798$\\
MF$_{tc}$ & $10.06$ & $8.032$ & $7.282$ & $6.741$ & $1.436$ & $3.066$ & $5.388$ & $8.101$ & $2.513$ & $4.438$ & $6.197$ & $7.359$ & $6.908$ & $1.790$\\
\midrule
HDMF & $\textbf{18.20}$& $\textbf{15.96}$ & $\textbf{13.61}$ & $\textbf{11.37}$ & $\textbf{5.510}$ & $\textbf{13.05}$ & $\textbf{21.13}$ & $\textbf{28.70}$ & $\textbf{8.458}$ & $\textbf{14.36}$ & $\textbf{16.56}$ & $\textbf{16.29}$ & $\textbf{11.50}$ & $\textbf{3.870}$ \\
\bottomrule
\end{tabular}\label{tab: performance}
\vspace{-1em}
\end{table*}
\end{small}
\vspace{-1em}

To ensure a fair comparison, the experiments are performed on the same real-world social-tagging dataset as used in~\cite{Xu2016,Zuo2016}, which is gathered from the Delicious bookmarking system
and released in HetRec 2011~\cite{Cantador:2011}. After using the same pre-processing to remove the infrequent tags that are used less than $15$ times, the resulting dataset is as shown in Table~\ref{tab:dataset}. We randomly select $80\%$ of assignments as training set, $5\%$ as validation set, and $15\%$ as test set.

All models are implemented using Python and Theano and run on a GPU server with an NVIDIA Tesla K$40$ GPU and $12$GB GPU memory. The parameters of HDMF are selected by grid search and the values are set as follows: (i) $\#$ of hidden layers is $5$; (ii)~$\#$ of neurons from $1$st to $5$th hidden layer are $2\,000$, $300$, $128$, $300$, and $2\,000$, respectively; (iii) the parameters $\lambda_\theta$ and $\lambda_e$ are set to $0.01$ and $0.2$; (iv)~the learning rate for model training is $0.002$.

In training, we first initialize the weight matrices $W_j$, using the random normal distribution, and initialize the biases~$b_j$ to be zero vectors; the model is then trained by back-propagation using stoch\-astic gradient descent; finally, the training stops when the model converges or reaches the maximum training runs. We also use the validation set to avoid over-fitting by early stopping.

As for the evaluation of recommendation systems, the most popular metrics are precision, recall, and F$1$-score~\cite{Bobadilla:2013}. Since users usually only browse the topmost recommended items, we apply these metrics at a given cut-off rank $k$, i.e., considering only the top-$k$ results on the recommendation list, called \textit{precision at $k$ ($P@k$)}, \textit{recall at $k$ ($R@k$)}, and \textit{F$1$-score at $k$ ($F@k$)}. In addition, since users always prefer to have their target items ranked in the front of the recommendation list, we also employ as evaluation metrics the \textit{mean average precision (MAP)} and the \textit{mean reciprocal rank (MRR)}, which take into account the order of items and give greater importance to the ones ranked higher.

\vspace{-0.4em}
\subsection{Results}

Table~\ref{tab: performance} depicts in detail the tag-aware personalized recommendation performances of our proposed HDMF and seven baselines on the Delicious dataset, in terms of $P@k$, $R@k$, $F@k$, MAP, and MRR, where four cut-off ranks $k=5$, $15$, $30$, and $50$ are selected.

In general, the relative performances of the baselines reported in Table~\ref{tab: performance} are highly consistent with the results reported in~\cite{Zuo2016}, \cite{Xu2016}, and \cite{Mano2016}; namely, (i) ACF outperforms CCF, (ii) DSPR outperforms CCF, ACF, and CCS, and (iii) MF$_{sf}$ and MF$_{ct}$ ``slightly'' outperform MF, respectively. More importantly, we note that our proposed model, HDMF, significantly outperforms all seven baselines in all metrics; e.g., the MRR (resp., MAP) of HDMF are $1.52$ (resp., $1.66$) times as high as that of the best baseline, DSPR (resp., MF$_{sf}$),  while the relative performances in \!$P@k$, \!$R@k$,~and \!$F@k$ are also similar. This finding strongly proves that by integrating the techniques of deep-semantic modeling, hybrid learning, and matrix factorization, HDMF overcomes the existing problems (as presented in Section~\ref{sec:Introduction}) of the state-of-the-art recommendation models and achieves very superior performance in tag-aware personalized recommendation.

Specifically, as shown in Table~\ref{tab: performance}, the MRR and MAP of HDMF are $1.52$ and $2.1$ times, respectively, as high as those of the-state-of-art deep-semantic model, DSPR. In addition, the relative improvements of HDMF to DSPR, in terms of \!$P@k$, \!$R@k$,~and \!$F@k$, all gradually enhance with the rise of the cut-off rank $k$, i.e., increasing from around $1.3$ times at $k=5$ to more than double at $k=50$. This observation demonstrates that incorporating collaborative-based capabilities (i.e., using correlation information between users and items to help the recommendation) can greatly enhance the deep-semantic model's performance in tag-aware recommendation, especially for the one with relative long recommendation lists.

Furthermore, by comparing the results of the matrix-factorization-based models, MF, MF$_{sf}$, and MF$_{tc}$, in Table~\ref{tab: performance}, we find that the AMF models, MF$_{sf}$ and~MF$_{tc}$, have close performances; and, more importantly, their relative improvements to MF are ``marginal'', e.g., their MAP and MRR are only $2.4\%$ and $6.8\%$, respectively, better than those of MF. This finding is actually consistent with the results in~\cite{Mano2016}, where the improvement rates of MF$_{sf}$ and~MF$_{tc}$ to MF are only $3.2\%$ and $5.5\%$, respectively. The reason for these ``marginal'' enhancements is as follows: the AMF models incorporate the additional source of information as a regularization term with a small coefficient in matrix factorization, which greatly limits the additional information's contribution on the optimizing gradient and thus limits their capabilities in improving the recommendation performance. By contrast, as shown in Table~\ref{tab: performance}, HDMF dramatically outperforms MF: the MAP and MRR of HDMF are about $70\%$ and $130\%$, respectively, better than those of MF. This is mainly because that the additional social tag information in HDMF is utilized to model the deep-semantic user and item matrices, which are then used directly as the decomposed user and item matrices in matrix factorization; since the decomposed matrices have dominant contribution on the optimizing gradient, HDMF maximizes the effect of the additional social tag information on the model optimization, making it possible to achieve significant improvements.



%



\vspace*{-1.5ex}

\section{Summary and Outlook}
\label{sec:Conclusion}

In this paper, we have briefly analyzed the state-of-the-art tag-aware personalized recommendation models that use content-based filtering or matrix factorization, and identified their existing problems. We thus have proposed a hybrid deep-semantic matrix factorization (HDMF) model to tackle these problems and to further enhance the performance of tag-aware personalized recommendation. We have also conducted extensive experimental studies and compared HDMF with seven state-of-the-art baselines; the results show that, by integrating the techniques of deep-semantic modeling, hybrid learning, and matrix factorization, HDMF greatly outperforms the state-of-the-art baselines in tag-aware personalized recommendation, in terms of all evaluation metrics.

In the future, further experiments will be conducted to compare the performances of HDMF on different kinds of Social Web datasets, e.g., Last.fm and MovieLens. Moreover, we will also investigate methodologies to add spatial and temporal information into the HDMF model to capture the users' real-time preferences.


\vspace*{-2.3ex}
{\small

}

\end{document}